\documentclass[12pt,a4paper]{article}
\usepackage{a4wide}
\usepackage{latexsym}
\usepackage{epsf}
\usepackage{amssymb}
\begin{document}
\def\be{\begin{equation}}
\def\ee{\end{equation}}
\def\bea{\begin{eqnarray}}
\def\eea{\end{eqnarray}}

\def\pd{\partial}
\def\a{\alpha}
\def\b{\beta}
\def\g{\gamma}
\def\d{\delta}
\def\m{\mu}
\def\n{\nu}
\newcommand{\fsl}{{\hspace{-7pt}\slash}}
\newcommand{\gsl}{{\hspace{-5pt}\slash}}
\newcommand{\dslash}{\pd\fsl}
\newcommand{\pslash}{p\gsl }
\newcommand{\Aslash}{A\gsl }
\newcommand{\qslash}{q\gsl }
\newcommand{\Lslash}{\Lambda\gsl }
\newcommand{\kuslash}{k_1\gsl }
\newcommand{\kdslash}{k_2\gsl }
\newcommand{\Dslash}{D\fsl}
\def \h{\mathcal{H}}
\def \hh{\mathcal{G}}
\def\t{\tau}
\def\p{\pi}
\def\th{\theta}
\def\l{\lambda}
\def\O{\Omega}
\def\r{\rho}
\def\s{\sigma}
\def\e{\epsilon}
  \def\scri{\mathcal{J}}
\def\cM{\mathcal{M}}
\def\tcM{\tilde{\mathcal{M}}}
\def\RR{\mathbb{R}}
%%%%%%%%%%%%%%%%%%%%%%%%%%%%%%%%%%%%%%%%%%%%%%%%%%%%%%%%%%%%%%%%%%%%%%

\hyphenation{re-pa-ra-me-tri-za-tion}
\hyphenation{trans-for-ma-tions}

%%%%%%%%%%%%%%%%%%%%%%%%%%%%%%%%%%%%%%%%%%%%%%%%%%%%%%%%%%%%%%%%%%%%%%

\begin{titlepage}

\begin{flushright}
IFT-UAM/CSIC-06-52\\
hep-ph/0610424\\
\end{flushright}

\vspace{1cm}

\begin{center}

{\bf\Large  Quantum corrections to Higher-Dimensional Theories.}

\vspace*{1.5cm}

{\bf Enrique \'Alvarez and Ant\'on F. Faedo }

\vspace{.3cm}

{\it  Instituto de F\'{\i}sica Te\'orica UAM/CSIC, C-XVI,\\
and\\  Departamento de F\'{\i}sica Te\'orica, C-XI,\\
  Universidad Aut\'onoma de Madrid 
  E-28049-Madrid, Spain }

\vskip 2cm

%%%%%%%%%%%%%%%%%%%%%%%%%%%%%%%%%%%%%%%%%%%%%%%%%%%%%%%%%%%%%%%%%%%%%%
%{\bf Abstract}
%\par

\begin{abstract}
This is a non-technical summary of the subtleties of quantum corrections on extra-dimensional theories:
should one first renormalize and then mode expand, or first expand in four-dimensional modes and then renormalize?
\end{abstract}

\end{center}

\end{titlepage}
%%%%%%%%%%%%%%%%%%%%%%%%%%%%%%%%%%%%%%%%%%%%%%%%%%%%%%%%%%%%%%%%%%%%%%

%\begin{quote}

%\end{quote}

%%%%%%%%%%%%%%%%%%%%%%%%%%%%%%%%%%%%%%%%%%%%%%%%%%%%%%%%%%%%%%%%%%%%%%

%\newpage
%%%%%%%%%%%%%%%%%%%%%%%%%%%%%%%%%%%%%%%%%%%%%%%%%%%%%%%%%%%%%%%%%%%%%%

%\setcounter{page}{1}
%\setcounter{footnote}{0}
%\renewcommand{\theequation}{\thesection.\arabic{equation}}

%\newpage
%%%%%%%%%%%%%%%%%%%%%%%%%%%%%%%%%%%%%%%%%%%%%%%%%%%%%%%%%%%%%%%%%%%%%%%%%

%%%%%%%%%%%%%%%%%%%%%%%%%%%%%%%%%%%%%%%%%%%%%%%%%%%%%%%%%%%%%%%%%%%%%%%%%%

In the past few years there has been an increasing interest in field theories 
defined in spacetimes of dimension greater than four. Such models, seen as low energy
effective theories of a more fundamental consistent theory like superstrings, provide a new
variety of very interesting mechanisms in order to solve long standing problems of the Standard Model.

Interesting possibilities are the idea of the Higgs particle originated from extra-dimensional 
components of gauge fields \cite{Manton}, often called Gauge-Higgs unification, and alternative 
mechanisms for symmetry breaking \cite{Hosotani, Scherk}. The best known of this kind of proposals 
are probably Large Extra Dimensions \cite{Hamed} and warped scenarios \cite{Randall}. 
These are only the original references, although the literature on the matter is very extensive.

A common problem in higher dimensional models is the neccesity to explain why extra dimensions
 are hidden, in the sense that the spacetime we experiment is effectively four-dimensional. Traditionally, 
extra dimensions are supposed to be compact and with a characteristic size extremely small\footnote{It is 
possible to avoid this requirement by using warped geometries with localized gravity \cite{Randall}} so that we
 would need energies unattainables in present colliders in order to directly detect them. Compactness of the
 extra dimensions allows us to expand fields propagating in the whole spacetime in harmonics and perform
 integrals over the extra coordenates. In that way we find a four-dimensional theory, but with an infinite
 number of fields corresponding to modes of the expansion: the so-called Kaluza-Klein modes.

We can then distinguish two viewpoints, the higher dimensional and the four-dimensional with the tower.
They are of course completely equivalent at the classical level. The question we are trying to answer is 
if this last statement remains true, and if so under what conditions, when one consider quantum corrections 
on both points of view. A negative result will be important because, as far as we know, the calculations
in the literature are almost always performed in four dimensions taking into account the tower 
(cf. however \cite{Garriga}). If the correct way to understand higher dimensional field
 theories is to compute radiative corrections directly in the complete spacetime (as we will try to argue), 
then a great number of results for the models considered before should be examined.

If we are interested in quantum effects it is sufficient to work to one loop order. To this order, 
the effective action is given in terms of a functional determinant 
\be\label{ea}
\Gamma\sim\log\,\det\,\Delta
\ee
where $\Delta$ is the operator representing the quadratic part of the action. In many interesting theories,
for instance the Standard Model, this last quantity is divergent. Extraction of the divergent part of (\ref{ea})
in a consistent way is the process of renormalization, in this case to one loop. There are several ways
of identifying the divergent part of (\ref{ea}), for example diagramatically in the sense of 't Hooft's
algorithm \cite{Hooft} generalized to the appropiate dimension\footnote{This algorithm is designed to 
give the poles in dimensional regularization in four dimensions, but it can be easily generalized to
an arbitrary {\it{even}} dimension. If one uses a proper time cutoff instead of dimensional regularization
it can also be applied to odd dimensions.}. A more effective approach, specially on curved backgrounds, 
is the heat kernel \cite{DeWitt}.

Concerning higher dimensional theories, it is then obvious that quantum equivalence requires the 
matching of the divergences on both points of view. The aim of this work is to explore whether this
 matching is possible or not.

It is important to say that in the particular case of a scalar interacting only through the universal 
coupling to an external gravitational field, after solving some subtleties, it is possible to perform
a clever resummation of the modes in a way that divergences coincide, although it is true that we find 
counterterms that we should not expect in a purely four-dimensional computation, as shown by Duff and Toms
in \cite{Duff}.

A crucial point in the argument is that the operators considered can be split into the form
\be
\Delta=\Delta_1+\Delta_2
\ee
where $\Delta_1$ acts trivially on the extra dimensional coordinates and $\Delta_2$ acts trivially on 
the usual four-dimensional ones\footnote{This requires a factorizable metric, so the gravitational background is not arbitrary}. 
Therefore, the result is not valid when the spliting does not take place, as happens on a 
warped background \footnote{A hint in that direction is given in \cite{Frolov}} as well as for a general
interacting theory.

We will focus our atention on a simple interacting theory, in particular Quantum Electrodynamics defined
on a six-dimensional manifold $\mathbb{R}^4\times S^1\times S^1$. The (Euclidean version of) the action is
\be\label{action}
S=\int d^6 x\left(\frac{1}{4}F_{MN}^2+\bar{\psi}(\Dslash+m)\psi\right)
\ee
An informed reader may notice that this action is non renormalizable, since the gauge coupling has mass dimension
$[e_6]=-1$. However, up to one loop this fact is not important in the sense that we can still identify and study
all the divergences. Before performing any calculation let us think a moment what should we expect to find
when one considers the one loop correction. 

As it is well known, the counterterms of the theory will be the most general six-dimensional operators compatible with the 
symmetries of the system, in this case a $U(1)$ gauge symmetry and Lorentz invariance. Thus the dimensionality
of the coupling allows us to write terms like
\be\label{ho}
e^2 D_MF^{MN}D^RF_{RN}\hspace{.8cm};\hspace{.8cm} e^2D_RF_{MN}D^RF^{MN}
\ee
Despite they were not present in the original Lagrangian, the radiative generation of these operators is 
unavoidable: the power of the coupling shows that is a one loop effect, they are of the right 
dimension and have the correct invariance. The appearance or this terms was discussed in \cite{Ghilencea, Oliver}.
Another important point is that the symmetry forbids a mass term for the gauge field, so the bosonic zero
modes remain massless at one loop. 

The explicit six-dimensional computation performed in \cite{Alvarez} agrees with these intuitions.

In order to perform a four-dimensional computation with the whole KK tower, let us expand the fields
in modes. Compactification of (\ref{action}) gives the four-dimensional gauge fixed action
\bea\label{tower}
\lefteqn{S=\int d^4 x\sum_{n_5,n_6}\left(\bar{\psi}^1_{n}\dslash \psi^1_n+\bar{\psi}^2_{n}
\dslash \psi^2_n+\bar{\psi}^1_{n}(i\frac{n_5}{R_5}+\frac{n_6}{R_6})\psi^2_n-
\bar{\psi}^2_{n}(i\frac{n_5}{R_5}-\frac{n_6}{R_6})\psi^1_n+\right.{}}\nonumber\\
&&{}+m\left(\bar{\psi}^1_{n}\psi^1_n-\bar{\psi}^2_{n}\psi^2_n\right)
-\frac{1}{2}(A_\m^n)^{*}\left(\Box-\frac{n_5^2}{R_5^2}-\frac{n_6^2}{R_6^2}\right) A^\m_n-
\frac{1}{2}(A_5^n)^{*}\left(\Box-\frac{n_5^2}{R_5^2}-\frac{n_6^2}{R_6^2}\right)A_5^n-
\nonumber\\
&&\left.-\frac{1}{2}(A_6^n)^{*}\left(\Box-\frac{n_5^2}{R_5^2}-
\frac{n_6^2}{R_6^2}\right)A_6^n-e\sum_m\left(\bar{\psi}^1_m\Aslash_{m-n}\psi^1_n+
\bar{\psi}^2_m\Aslash_{m-n}\psi^2_n
+\right.\right.{}\nonumber\\&&{}\left.+\bar{\psi}^1_mA_5^{m-n}\psi^2_n-
\bar{\psi}^2_mA_5^{m-n}\psi^1_n-i\bar{\psi}^1_mA_6^{m-n}\psi^2_n-
i\bar{\psi}^2_mA_6^{m-n}\psi^1_n\right)\Bigg)
\eea
One has to double the number of fermions because in $d$ dimensions they have $2^{[d/2]}$ components
(eight in six dimensions, four in four dimensions). Also the extra components of the gauge field $A_5^n$ and $A_6^n$
appear as four-dimensional scalars\footnote{They are identified with the Higgs in gauge-Higgs unification.}. It 
is important to note that the spacetime symmetry is spontaneously broken to
\be
O(6)\longrightarrow O(4)\times O(2)\times O(2)
\ee
While the extra-dimensional gauge symmetry traduces into the infinte set of four-dimensional symmetries
\bea
\d A_\m^n&=&i\pd_\m \Lambda_n\nonumber\\
\d A_5^n&=&-\frac{n_5}{R_5} \Lambda_n\nonumber\\
\d A_6^n&=&-\frac{n_6}{R_6} \Lambda_n
\eea
Please note that the scalar zero modes are singlets under a gauge transformation.
Finally the coupling is now dimensionless, as it is defined by
\be
e\equiv \frac{e_6}{\sqrt{R_5 R_6}}\equiv e_6M
\ee
Let us repeat the exercise done with the previous action and ask ourselves what kind of corrections
one would expect. First of all, the coupling is dimensionless so we cannot use it to reduce the dimension
of higher order operators. Therefore, terms like the ones in (\ref{ho}) are forbidden, at least in perturbation theory.

Next, since the scalar zero mode is singlet there are no symmetries to protect its mass against radiative
corrections, as it happens with the Standard Model Higgs. Then we expect a mass term for it (in fact there
is no reason not to expect operators of higher power, i.e. cubic or quartic interactions).

Another important point is that the gauge zero mode $A_\m^0$, which plays the role of the usual photon,
couples diagonally to an infinite tower of fermions, with the same strengh as in four-dimensional QED and 
to every fermion. The only diference betwen the fermions of the tower is their masses,
 which are labelled by a pair of integers. Now, the pole in the Vacuum Polarization Function does not depend
on the mass of the fermion running in the loop. We should have then the same contribution to the
$\b$-function as in QED for every fermion. Since the number of fermions is infinite, one has to sum the same quantity 
an infinite number of times. This gives rise to an additional divergence coming from the sum. One can think
that this is the expected effect of an infinite number of fields interacting all to each other. 

Again all these expectations are confirmed with standard computatios and the explicit result can be found in \cite{Alvarez}.
Of course it seems impossible to reconcile both points of view.
A natural question is to what extent this is the consequence of the non renormalizability of the model. Unfortunatelly
studies along these lines but with a renormalizable theory (in particular four dimensional $QED$) show that the inequivalence
has nothing to do with renormalizability.

In fact, the case of $QED_4$ on the four-dimensional manifold $\mathbb{R}^2\times S^1\times S^1$ provides the most transparent example
of this kind of effects, so it is worth to study it. 

The counterterm calculated in the whole spacetime is the usual one of QED, which yields the well known $\b$-function
\be
\b=\frac{e^3}{12\p^2}
\ee
Or in terms of the two-dimensional coupling $\bar{e}=eM$
\be\label{run}
\bar{e}^2=\frac{\bar{e}_0^2}{1-\frac{\bar{e}_0^2}{6\p^2M^2}\log{\frac{\m}{\m_0}}}
\ee
On the other hand, symmetry forbids again a mass term form the gauge boson. Moreover, in four dimensions
even if we include explicitly a mass term for the gauge boson in the bare Lagrangian its mass does not receive 
radiative corrections and remains unrenormalized \cite{Collins}.

From a two-dimensional perspective the situation is radically different. The superficial degree of divergence
of a diagram is now
\be
D=2-\frac{1}{2}E_f-V
\ee
where $V$ is the number of vertices and $E_f$ is the number of fermionic external lines. This means that 
any diagram with fermions in external lines can be primitivelly divergent. Thus, there are no counterterms
for the fermionic sector, a fact that is impossible to justify thinking in four dimensions. The primitively
divergent diagrams involve only bosons as external states. 

Morover, the Vacuum Polarization Function is known to be finite in two dimensions (remember the Schwinger model).
This means that the only divergent correction, apart from a tadpole, is the two point function of the 
two-dimensional scalars $A_3^n$ and $A_4^n$. The zero mode was massless at tree level, but now since it is a 
gauge singlet it gets mass through radiative corrections. This was impossible in four dimensions as we have said.
Also there is no running of the coupling at all, in clear contradiction with (\ref{run}), although possible 
deviations from $e^2\approx e_0^2$ can be seen only in energies exponential in the compactification mass 
\be
\frac{\m}{\m_0}\gg e^{\frac{6\p^2M^2}{\bar{e}_0^2}}
\ee
The explicit counterterm is given in \cite{Alvarez} but its properties are basically the ones explained here. 

The conclusion is that there is some sort of quantum inequivalence between Kaluza-Klein models
when one considers loop corrections in the whole spacetime or in the dimensionally reduced theory.
Therefore one has to take care when computing in extra dimensional field theories, at least when dealing
with radiative corrections. In that case, since one considers effects at energies much higher than
the compactification scale, the compact dimensions should be seen in the same way as the usual four and
the spacetime should be higher-dimensional. The natural way of computing is then in the whole manifold
performing next the mode expansion to get four-dimensional quantities. A more detailed argumentation of this way
of thinking can be found in \cite{Alvarez2}.

%%%%%%%%%%%%%%%%%%%%%%%%%%%%%%%%%%%%%%%%%%%%%%%%%%%%%%%%%%%%%%%%%%%%%%%%%%%%%%%%%%%%%%%%%%%%%%
\section*{Acknowledgments}
%%%%%%%%%%%%%%%%%%%%%%%%%%%%%%%%%%%%%%%%%%%%%%%%%%%%%%%%%%%%%%%%%%%%%%%%%%%%%%%%%%%%%%%%%%%%%%%

This work has been partially supported by the European Commission (HPRN-CT-200-00148) and by FPA2003-04597 (DGI del MCyT, Spain), as well as Proyecto HEPHACOS (CAM); P-ESP-00346. A.F. Faedo has been supported by a MEC grant, AP-2004-0921. We are indebted to Bel\'en Gavela, C\'esar G\'omez and Karl Landsteiner for useful discussions and to D.J.Toms for illuminating correspondence. We would also like to thank Joan Sol\'a for kindly inviting us to the Conference.

%%%%%%%%%%%%%%%%%%%%%%%%%%%%%%%%%%%%%%%%%%%%%%
               

\begin{thebibliography}{99}
%%%%%%%%%%%%%%%%%%%%%%%%%%%%%%%%%%%%%%%%%%%%%%


%\cite{Alvarez:2006we}
\bibitem{Alvarez}
  E.~Alvarez and A.~F.~Faedo,
  ``Renormalized Kaluza-Klein theories,''
  JHEP {\bf 0605} (2006) 046
  [arXiv:hep-th/0602150].
  %%CITATION = HEP-TH 0602150;%%



%\cite{Alvarez:2006sf}
\bibitem{Alvarez2}
  E.~Alvarez and A.~F.~Faedo,
  ``Renormalized masses of heavy Kaluza-Klein states,''
  arXiv:hep-th/0606267.
  %%CITATION = HEP-TH 0606267;%%



%\cite{Arkani-Hamed:1998rs}
\bibitem{Hamed}
  N.~Arkani-Hamed, S.~Dimopoulos and G.~R.~Dvali,
  ``The hierarchy problem and new dimensions at a millimeter,''
  Phys.\ Lett.\ B {\bf 429} (1998) 263
  [arXiv:hep-ph/9803315].
  %%CITATION = HEP-PH 9803315;%%
  I.~Antoniadis, N.~Arkani-Hamed, S.~Dimopoulos and G.~R.~Dvali,
  ``New dimensions at a millimeter to a Fermi and superstrings at a TeV,''
  Phys.\ Lett.\ B {\bf 436} (1998) 257
  [arXiv:hep-ph/9804398].
  %%CITATION = HEP-PH 9804398;%%


%\cite{Collins:1984xc}
\bibitem{Collins}
  J.~C.~Collins,
   ``Renormalization. An Introduction To Renormalization, The Renormalization Group, And The Operator Product Expansion,''
%\href{http://www.slac.stanford.edu/spires/find/hep/www?irn=1341391}{SPIRES entry}




%\cite{DeWitt:1988fm}
\bibitem{DeWitt}
  B.~S.~DeWitt,
  ``Dynamical Theory of Groups and Fields''
%\href{http://www.slac.stanford.edu/spires/find/hep/www?irn=1930923}{SPIRES entry}



%\cite{Duff:1982wm}
\bibitem{Duff}
  M.~J.~Duff and D.~J.~Toms,
  ``Divergences And Anomalies In Kaluza-Klein Theories,''
CERN-TH-3248
%\href{http://www.slac.stanford.edu/spires/find/hep/www?r=cern-th-3248}{SPIRES entry}
{\it Presented at Second Seminar on Quantum Gravity, Moscow, USSR, Oct 13-15, 1981}
M.~J.~Duff and D.~J.~Toms,
  ``Kaluza-Klein Kounterterms,''
CERN-TH-3259
%\href{http://www.slac.stanford.edu/spires/find/hep/www?r=cern-th-3259}{SPIRES entry}
{\it Presented at 2nd Europhysics Study Conf. on Unification of Fundamental Interactions, Erice, Sicily, Oct 6-14, 1981}


%\cite{Frolov:1999an}
\bibitem{Frolov}
  V.~P.~Frolov, P.~Sutton and A.~Zelnikov,
  ``The dimensional-reduction anomaly,''
  Phys.\ Rev.\ D {\bf 61} (2000) 024021
  [arXiv:hep-th/9909086].
  %%CITATION = HEP-TH 9909086;%%


%\cite{Garriga:2001ar}
\bibitem{Garriga}
  J.~Garriga, O.~Pujolas and T.~Tanaka,
  ``Moduli effective potential in warped-brane world compactifications,''
  Nucl.\ Phys.\ B {\bf 655} (2003) 127
  [arXiv:hep-th/0111277].
  %%CITATION = HEP-TH 0111277;%%



%\cite{Ghilencea:2004sq}
\bibitem{Ghilencea}
  D.~M.~Ghilencea,
  ``Higher derivative operators as loop counterterms in one-dimensional field theory orbifolds,''
  JHEP {\bf 0503} (2005) 009
  [arXiv:hep-ph/0409214].
  %%CITATION = HEP-PH 0409214;%%



%\cite{Hosotani:1983xw}
\bibitem{Hosotani}
  Y.~Hosotani,
  ``Dynamical Mass Generation By Compact Extra Dimensions,''
  Phys.\ Lett.\ B {\bf 126} (1983) 309.
  %%CITATION = PHLTA,B126,309;%%
  ``Dynamics of Nonintegrable Phases and Gauge Symmetry Breaking,''
  Annals Phys.\  {\bf 190} (1989) 233.
  %%CITATION = APNYA,190,233;%%


%\cite{Manton:1979kb}
\bibitem{Manton}
  N.~S.~Manton,
  ``A New Six-Dimensional Approach To The Weinberg-Salam Model,''
  Nucl.\ Phys.\ B {\bf 158} (1979) 141.
  %%CITATION = NUPHA,B158,141;%%
  D.~B.~Fairlie,
  ``Higgs Fields And The Determination Of The Weinberg Angle,''
  Phys.\ Lett.\ B {\bf 82} (1979) 97.
  %%CITATION = PHLTA,B82,97;%%
  P.~Forgacs and N.~S.~Manton,
  ``Space-Time Symmetries In Gauge Theories,''
  Commun.\ Math.\ Phys.\  {\bf 72} (1980) 15.
  %%CITATION = CMPHA,72,15;%%
  S.~Randjbar-Daemi, A.~Salam and J.~A.~Strathdee,
  ``Spontaneous Compactification In Six-Dimensional Einstein-Maxwell Theory,''
  Nucl.\ Phys.\ B {\bf 214} (1983) 491.
  %%CITATION = NUPHA,B214,491;%%


%\cite{Oliver:2003cy}
\bibitem{Oliver}
  J.~F.~Oliver, J.~Papavassiliou and A.~Santamaria,
  ``Can power corrections be reliably computed in models with extra dimensions?,''
  Phys.\ Rev.\ D {\bf 67} (2003) 125004
  [arXiv:hep-ph/0302083].
  %%CITATION = HEP-PH 0302083;%%



%\cite{Ramond:1989yd}
%\bibitem{Ramond}
%  P.~Ramond,
%  ``Field Theory: a modern primer,''
%  Front.\ Phys.\  {\bf 74} (1989) 1.
  %%CITATION = FRPHA,74,1;%%




%\cite{Randall:1999vf}
\bibitem{Randall}
L.~Randall and R.~Sundrum,
``An alternative to compactification,''
Phys.\ Rev.\ Lett.\  {\bf 83} (1999) 4690
[arXiv:hep-th/9906064].\\
%%CITATION = HEP-TH 9906064;%%
``A large mass hierarchy from a small extra dimension,''
Phys.\ Rev.\ Lett.\  {\bf 83} (1999) 3370
[arXiv:hep-ph/9905221].
%%CITATION = HEP-PH 9905221;%%



%\cite{Scherk:1978ta}
\bibitem{Scherk}
  J.~Scherk and J.~H.~Schwarz,
  ``Spontaneous Breaking Of Supersymmetry Through Dimensional Reduction,''
  Phys.\ Lett.\ B {\bf 82} (1979) 60.
  %%CITATION = PHLTA,B82,60;%%
  E.~Cremmer, J.~Scherk and J.~H.~Schwarz,
  ``Spontaneously Broken N=8 Supergravity,''
  Phys.\ Lett.\ B {\bf 84} (1979) 83.
  %%CITATION = PHLTA,B84,83;%%


%\cite{'tHooft:1973us}
\bibitem{Hooft}
  G.~'t Hooft,
   ``An algorithm for the poles at dimension four in the dimensional regularization procedure,''
  Nucl.\ Phys.\ B {\bf 62} (1973) 444.
  %%CITATION = NUPHA,B62,444;%%



%\cite{Vassilevich:2003xt}
%\bibitem{Vassilevich}
%  D.~V.~Vassilevich,
%  ``Heat kernel expansion: User's manual,''
%  Phys.\ Rept.\  {\bf 388} (2003) 279
%  [arXiv:hep-th/0306138].
%%CITATION = HEP-TH 0306138;%%

              
%%%%%%%%%%%%%%%%%%%%%%%%%%%%%%%%%%%%%%%%%%%%%%%%%%%%%%%%%%%
\end{thebibliography}
\end{document}